\documentclass[floatfix,prd, nofootinbib, twocolumn, preprintnumbers, superscriptaddress]{revtex4-1}
\usepackage[utf8]{inputenc}
\usepackage{mathrsfs,bm}
\usepackage{natbib}
\usepackage{txfonts}
\usepackage{soul}
\usepackage{ulem}
\usepackage{amssymb}
\usepackage{rotating}
\usepackage{indentfirst}
\usepackage{graphicx,booktabs}
\usepackage{multirow}
\usepackage{slashed}
\usepackage{overpic}
\usepackage{color}
\usepackage{hyperref}
\usepackage{bbding}
\usepackage{url}

\begin{document}
\title{Are we close to solving the puzzle of weak radiative hyperon decays?}

\author{Rui-Xiang Shi}
\affiliation{School of Space and Environment,  Beihang University, Beijing 102206, China}
\affiliation{School of Physics, Beihang University, Beijing 102206, China}



\author{Li-Sheng Geng}
\email[Corresponding author: ]{lisheng.geng@buaa.edu.cn}
\affiliation{School of
Physics,  Beihang University, Beijing 102206, China}
\affiliation{Peng Huanwu Collaborative Center for Research and Education, Beihang University, Beijing 100191, China}
\affiliation{Beijing Key Laboratory of Advanced Nuclear Materials and Physics, Beihang University, Beijing 100191, China}

\begin{abstract}
The BESIII experiment at the Beijing Electron Positron collider,  run as a
 ``hyperon factory'', recently reported the first measurement of the asymmetry parameter of the $\Lambda\to n\gamma$ decay and updated its branching fraction, which differs from the PDG average by 5.6 sigma. We highlight the impact of this new measurement on our understanding of the  puzzle of  weak radiative hyperon decays and pinpoint what needs to be done in the future. 
\end{abstract}

\maketitle

Hyperons, which carry odd half-integer spins and nonzero baryon number, are baryons containing  one or more strange quarks, but no charm, bottom, or top quark. They are predominantly produced via the strong interaction while decay through the weak interaction.  Weak decays of hyperons can be divided into three categories: non-leptonic, semi-leptonic, and radiative decays. Among them, weak radiative hyperon decays~(WRHDs) are unique physical processes involving simultaneously the strong, weak, and electromagnetic interactions, which are generically denoted as $s\to d\gamma$ at the quark level. In general, the electroweak interaction is  well understood. Therefore, studies of WRHDs are important to test the strong interaction and search for beyond the standard model physics.

Hereafter we focus on the weak radiative decays of hyperons of the lowest-lying  spin-1/2 baryon octet, which contain six  decay channels, i.e., $\Lambda\to n\gamma$, $\Sigma^+\to p\gamma$, $\Sigma^0\to n\gamma$, $\Xi^0\to\Lambda\gamma$, $\Xi^0\to\Sigma^0\gamma$, and $\Xi^-\to\Sigma^-\gamma$. There may be other weak radiative decays of  spin-3/2 decuplet baryons, for instance, $\Omega^-\to\Xi^-\gamma$. However, we do not discuss them here because their experimental measurements  are currently not yet available~\cite{Li:2016tlt}. In the future, the $\Omega^-\to\Xi^-\gamma$ decay may be measured via the $\psi(2S)\to\Omega^-\bar{\Omega}^+$ decay at the proposed super tau-charm factory (STCF)~\cite{Zhou:2021rgi,Barnyakov:2020vob}.

The effective Lagrangian for the weak radiative decay of a hyperon $B_i$ of momentum $p$ to a baryon $B_f$
of momentum $k$ and a photon of momentum $p-k$ reads
\begin{eqnarray}
{\cal L}=\frac{eG_F}{2}\bar{B}_f(p)(a+b\gamma_5)\sigma^{\mu\nu}B_i(k)F_{\mu\nu},\label{WRHDs:lag}
\end{eqnarray}
where $G_F$ is the Fermi constant, $e$ is the electron charge and $F_{\mu\nu}=\partial_\mu A_\nu-\partial_\nu A_\mu$ where $A_\mu$ is the photon field. $\sigma^{\mu\nu}$ and $\gamma_5$ are the combinations of  Dirac gamma matrices. The amplitudes $a$ and $b$ with dimension of masses are partity-conserving and partiy-violating, respectively.

For a polarized spin-1/2 hyperon, the angular distribution of its radiative decay is
\begin{eqnarray}
&&\frac{d\Gamma}{d\cos\theta}=\frac{e^2G_F^2}{\pi}(|a|^2+|b|^2)[1+\frac{2{\rm Re}(ab^*)}{|a|^2+|b|^2}\cos\theta]\cdot|\vec{k}|^3,\nonumber\\
&&\alpha_\gamma=\frac{2{\rm Re}(ab^*)}{|a|^2+|b|^2},\label{WRHDs:asypara}
\end{eqnarray}
where $\alpha_\gamma$ is the asymmetry parameter, reflecting the mixing of parity-conserving and parity-violating processes, $\theta$ is the angle between the spin of the initial hyperon $B_i$ and the 3-momentum $|\vec{k}|=\frac{m_i^2-m_f^2}{2m_i}$ of the final baryon $B_f$. It is clear that non-zero asymmetry parameters require non-vanishing  amplitudes $a$ and $b$. 

In 1964, Yasuo Hara, a theoretical physicist, proposed the renowned Hara theorem~\cite{Hara:1964zz} which is crucial to understand the theoretical implications of WRHDs. This theorem is based on three general symmetry considerations: gauge invariance, CP conservation, and U-spin symmetry (which is a subset of SU(3) flavor symmetry), i.e., interchange of $s$ and $d$ quarks. To illustrate Hara's theorem, let us take the $\Sigma^+\to p\gamma$ decay as an example. By considering gauge invariance and CP symmetry, one has the most general form of the effective Lagrangian ${\cal L}_{\rm p.c.}+{\cal L}_{\rm p.v.}$, which read
\begin{eqnarray}
&&{\cal L}_{\rm p.c.}=a\left(\bar{p}\sigma^{\mu\nu}\Sigma^+F_{\mu\nu}+
\bar{\Sigma}^+\sigma^{\mu\nu}pF_{\mu\nu}\right)\frac{eG_F}{2},\\
&&{\cal L}_{\rm p.v.}=b\left(\bar{p}\sigma^{\mu\nu}\gamma_5\Sigma^+F_{\mu\nu}-
\bar{\Sigma}^+\sigma^{\mu\nu}\gamma_5pF_{\mu\nu}\right)\frac{eG_F}{2},
\end{eqnarray}
where the parity-conserving term ${\cal L}_{\rm p.c.}$ is symmetric under the exchange of $\Sigma^+$ and $p$  while the parity-violating term ${\cal L}_{\rm p.v.}$ is antisymmetric. In a U-spin symmetric world where one can exchange $\Sigma^+$ with $p$, the effective Lagrangian before and after U-spin rotation is identical, which leads to the constraint $b=-b$, i.e., $b=0$. Thus Hara's theorem dictates that the
parity-violating decay amplitude $b$ for the $\Sigma^+\to p\gamma$ decay vanishes in the U-spin symmetry limit. In other words, the asymmetry parameter for the $\Sigma^+\to p\gamma$ decay should not be too large even taking into account the effect of U-spin symmetry breaking. Similar conclusions can be obtained for the $\Xi^-\to\Sigma^-\gamma$ decay.

However, in 1969 the first experimental measurement of the $\Sigma^+\to p\gamma$ decay yielded a surprisingly large asymmetry parameter~\cite{Gershwin:1969fpe}. This experimental result is consistent with the most recent Review of Particle Physics (RPP)~\cite{ParticleDataGroup:2022pth} and BESIII data~\cite{BESIII:2023fhs}, which contradicts Hara’s theorem. Such phenomena are referred to as  the puzzle of WRHDs. For over fifty years, theorists have  struggled
to understand the puzzling result. A variety of phenomenological models~\cite{Gavela:1980bp,Nardulli:1987ub,Zenczykowski:1991mx,Zenczykowski:2005cs,Dubovik:2008zz,Niu:2020aoz} were proposed with or without evading Hara's theorem and described well the $\Sigma^+\to p\gamma$ data. However, the puzzle of WRHDs became more complicated with time. That is, the predictions of those models cannot well accommodate the other WRHDs, i.e., $\Xi^0\to\Lambda\gamma$ and $\Xi^0\to\Sigma^0\gamma$.  A more detailed discussion can be found in Ref.~\cite{Shi:2022dhw}. In particular the baryon chiral perturbation theory~($\chi$PT)~\cite{Jenkins:1992ab,Neufeld:1992np,Borasoy:1999nt} as the low-energy effective field theory~(EFT) of quantum chromodynamics (QCD) also failed to provide a unified picture for all the WRHDs, which satisfies Hara's theorem.

It is worth noting that the current values~\cite{ParticleDataGroup:2022pth} of the asymmetry parameters of the $\Xi^0\to\Lambda\gamma$ and $\Xi^0\to\Sigma^0\gamma$ decays have changed dramatically because their signs were wrongly assigned in the first measurements~(see Ref.~\cite{Shi:2022dhw} and references cited therein). Furthermore, the $\Sigma^0\to n\gamma$ decay, which is overwhelmed by the electromagnetic $\Sigma^0\to\Lambda\gamma$ decay, has not been measured experimentally. In addition,  the $\Lambda\to n\gamma$ decay presents severe problems on the experimental side. Specifically, previous fixed target experiments either lacked enough statistics or failed to produce polarized $\Lambda$ hyperons. Thus the asymmetry parameter of $\Lambda\to n\gamma$ is unknown till recently.

The Beijing Electron Spectrometer III~(BESIII) experiment operating
at the Beijing Electron-Positron Collider II~(BEPCII) provides an ideal environment to study  hyperon decays, which has collected $\sim10^{10}$ $e^+e^-\to J/\psi$ events~\cite{Li:2016tlt}. A substantial amount of spin-1/2 hyperon pairs $B\bar{B}$ can be produced in $J/\psi\to B\bar{B}$ decays. Therefore, the BESIII experiment can be run as a ``hyperon factory'' in the characteristically clean environment of an electron-positron collider. In addition, one can use the mature ``tag technique'' to fully reconstruct one of the hyperons and look at the recoil side in the $J/\psi$ decay into hyperon pairs. For several years, this technique has led to  remarkable achievements in the measurements of  hyperon non-leptonic decays~\cite{BESIII:2021ypr,BESIII:2018cnd,BESIII:2022qax}, which play a vital role in model-independent studies of  WRHDs~\cite{Jenkins:1992ab,Neufeld:1992np,Borasoy:1999nt}.

\begin{figure*}[htb]
  \centering
  \includegraphics[width=0.9\linewidth]{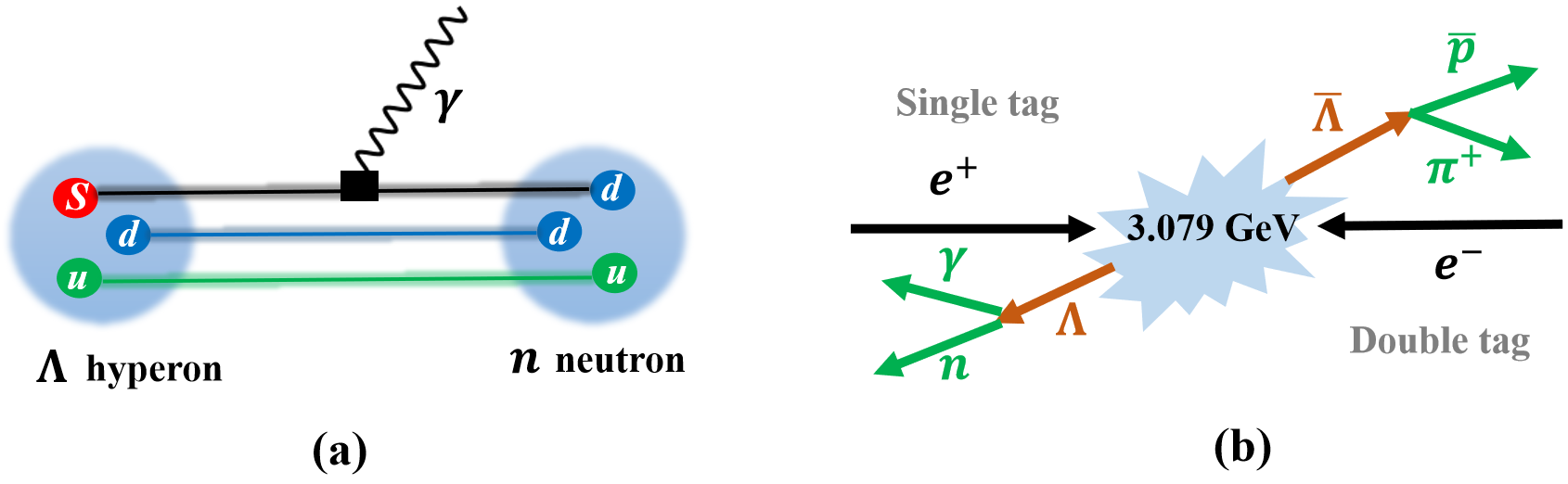}\\
  \caption{(Color online) Left panel: Feynman diagram for the $\Lambda\to n\gamma$ decay, which is $s\to d\gamma$  at the quark level. Right panel: Schematic diagram for the measurement of the $\Lambda\to n\gamma$ decay in the BESIII experiment.}\label{Fig1}
\end{figure*}

The BESIII Collaboration has started a campain of measuring  the $\Lambda\to n\gamma$ decay~(Fig.~\ref{Fig1}~a) by collecting hyperon-antihyperon resonant pair production at the $J/\psi$ peak. By utilizing a double-tag technique, the signal events of $\Lambda\to n\gamma$ in the BESIII experiment can be extracted from the cascade decay $J/\psi\to\Lambda\bar{\Lambda}$ with $\Lambda\to n\gamma$ and $\bar{\Lambda}\to\pi^+\bar{p}$. The center-of-mass energy is set at 3.079~{\rm GeV}. As shown in Fig.~\ref{Fig1}~b, the $\Lambda$ hyperon is reconstructed by the neutral photon $\gamma$ and neutron $n$, denoted as double tag. In the selection of a single tag event, the decay reconstructed is $\bar{\Lambda}\to\pi^+\bar{p}$. The asymmetry 
parameter $\alpha_\gamma$ for the signal channel $\Lambda\to n\gamma$ can be 
determined via the angular distribution 
of the final-state baryon. In the decay of a  polarized hyperon $\Lambda$ into $n\gamma$, the angular distribution of the neutron momentum in
the rest frame of the $\Lambda$ hyperon is given as
\begin{eqnarray}
\frac{dN}{d\Omega}\propto1+\alpha_\gamma {\bf P}_\Lambda\cdot\hat{\bf n},
\end{eqnarray}
where ${\bf P}_\Lambda$ is the polarization vector of the $\Lambda$ hyperon, and $\hat{\bf n}$ is the unit vector of the neutron momentum in the rest frame of the initial particle $\Lambda$. Compared with previous fixed target experiments, the BESIII experiment has the following advantages~\cite{BESIII:2022rgl}: (i) The total number of hyperons can be
determined by the double-tag technique and therefore it can achieve the direct measurement of an absolute branching fraction for the $\Lambda\to n\gamma$ decay with clean background. (ii) The decay  $J/\psi\to\Lambda\bar{\Lambda}$ produced a large number of polarized $\Lambda$ hyperons and the angular distribution of final states contains the $\alpha_\gamma$ information. (iii) The $J/\psi$ inclusive decay channels can provide control samples needed in data analysis and reduce systematic uncertainties in the measurement. 

\begin{figure}[h!]
  \centering
  \includegraphics[width=0.9\linewidth]{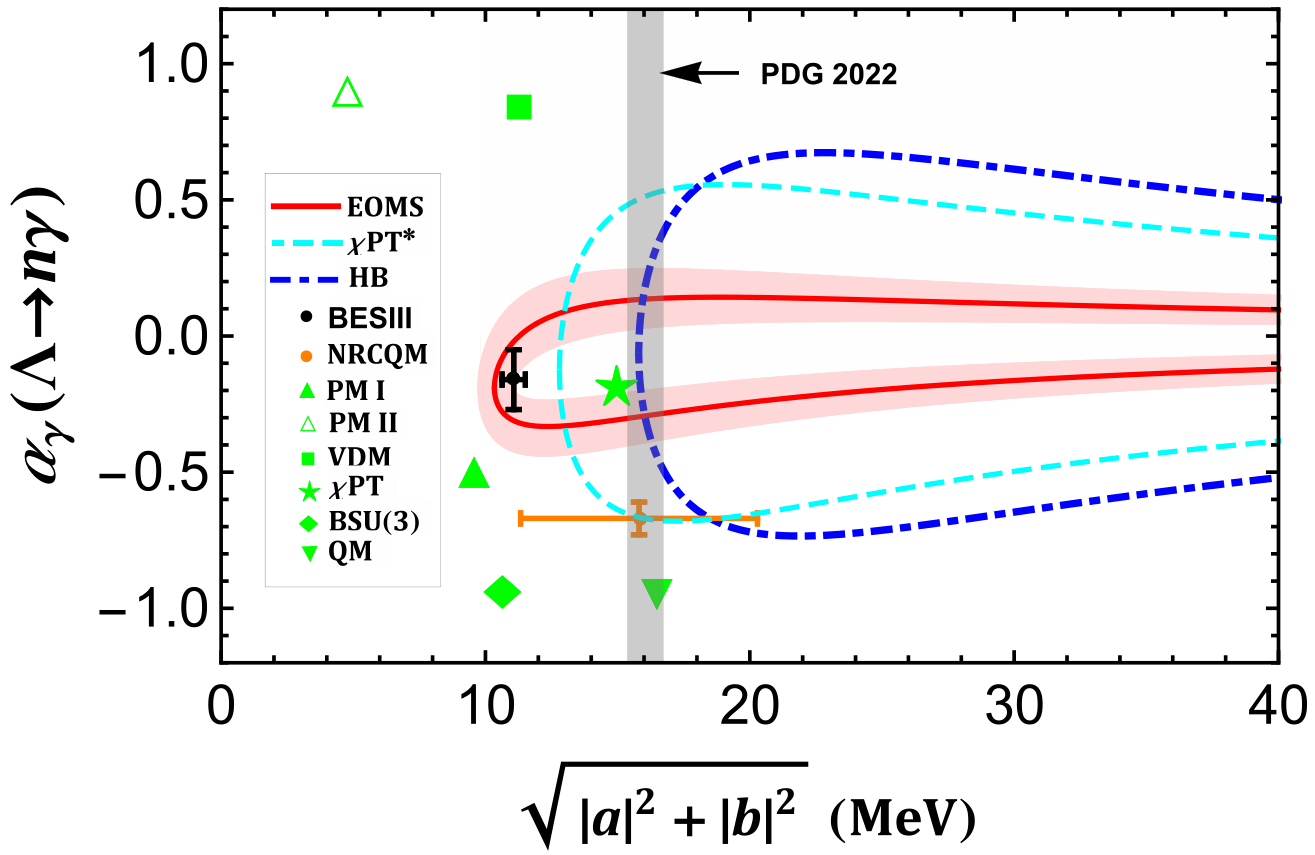}\\
  \caption{(Color online) Asymmetry parameter $\alpha_\gamma$ of  the $\Lambda\to n\gamma$ decay as a function of $\sqrt{|a|^2+|b|^2}$. The solid point in black with $xy$errorbars corresponds to the new BESIII data~\cite{BESIII:2022rgl} and that in orange is the prediction of the non-relativistic constituent quark model~(NRCQM)~\cite{Niu:2020aoz}. The gray band represents the current PDG average~\cite{ParticleDataGroup:2022pth}. The solid line and band in red denote the results of the EOMS $\chi{\rm PT}$~\cite{Shi:2022dhw}. The blue dot-dashed and cyan dashed lines correspond to the heavy baryon (HB) $\chi{\rm PT}$ results~\cite{Jenkins:1992ab} and $\chi{\rm PT}$ predictions~\cite{Neufeld:1992np} in the covariant formulation without the counter-term contribution to amplitude $b$. Other symbols in green stand for the results predicted in the pole models~(solid Delta: PM I~\cite{Gavela:1980bp} and hollow Delta: PM II~\cite{Nardulli:1987ub}), vector-dominance model~(solid square: VDM)~\cite{Zenczykowski:1991mx}, chiral perturbation theory at tree level~(solid star: $\chi{\rm PT}$)~\cite{Borasoy:1999nt}, broken SU(3) model~(solid rhombus: BSU(3))~\cite{Zenczykowski:2005cs}, and quark model~(solid nabla: QM)~\cite{Dubovik:2008zz}, respectively.}\label{Fig2}
\end{figure}

In June 2022, the BESIII Collaboration reported the first absolute branching fraction and asymmetry parameter of $\Lambda\to n\gamma$~\cite{BESIII:2022rgl}. For the purpose of understanding the experimental result,  the asymmetry parameter $\alpha_\gamma$ for the $\Lambda\to n\gamma$ decay is plotted in Fig.~\ref{Fig2} as a function of $\sqrt{|a|^2+|b|^2}$, which can be viewed as the branching fraction.   Surprisingly, the measured branching fraction deviates from the current PDG average~\cite{ParticleDataGroup:2022pth}~(band in gray) with a significance of $5.6\sigma$. None of the previous predictions in the pole models~(PM)~\cite{Gavela:1980bp,Nardulli:1987ub}, the vector-dominance model~(VDM)~\cite{Zenczykowski:1991mx}, the chiral perturbation theory~($\chi$PT)~\cite{Jenkins:1992ab,Neufeld:1992np,Borasoy:1999nt}, the broken SU(3) model~\cite{Zenczykowski:2005cs}, the quark model~(QM)~\cite{Dubovik:2008zz} and the non-relativistic constituent quark model~(NRCQM)~\cite{Niu:2020aoz} can describe the new data. In this sense, the BESIII results further exacerbate the puzzle of WRHDs.

After the BESIII Collaboration released the $\Lambda\to n\gamma$ results, some of us reported  a comprehensive study of all the WRHDs in covariant baryon chiral perturbation theory with the extended-on-mass-shell~(EOMS) renormalization scheme~\cite{Shi:2022dhw}. Compared with the previous covariant $\chi$PT study~\cite{Neufeld:1992np}, the latest study~\cite{Shi:2022dhw} in the EOMS $\chi$PT  updated the  values of the relevant low energy constants, ensured a consistent power counting, and determined the counter-term contributions for the first time. One finds that the predictions of the EOMS $\chi$PT for the $\Lambda\to n\gamma$ decay are in good agreement with the new BESIII measurements, as shown by the red band in Fig.~\ref{Fig2}. Furthermore, the EOMS $\chi$PT can describe simultaneously the $\Xi^0\to\Sigma^0\gamma$, $\Xi^0\to\Lambda^0\gamma$ and $\Xi^-\to\Sigma^-\gamma$ decays as well. However,  for the asymmetry parameter of the $\Sigma^+\to p\gamma$ decay, there is still a sizable  difference between the predictions of the EOMS $\chi$PT and the current experimental data~\cite{ParticleDataGroup:2022pth}. 

In Table~\ref{tab1} we compare the results of the broken SU(3) model~\cite{Zenczykowski:2005cs}, the QM~\cite{Dubovik:2008zz} and the NRCQM~\cite{Niu:2020aoz} with the experimental data. Note that these works~\cite{Zenczykowski:2005cs,Dubovik:2008zz,Niu:2020aoz} are published after the last major update of experimental data except for the latest BESIII measurement. Somehow surprisingly, the NRCQM  predicted wrong signs for the asymmetry parameters of the $\Xi^0\to \Lambda/\Sigma^0 \gamma$ decay modes, while both the QM and the broken SU(3) model reproduced pretty well these asymmetry parameters.  We note that the NRCQM did not fit the data of WRHDs but took the parameters of their model from spectroscopy studies~\cite{Niu:2020aoz}.  On the other hand, both the broken SU(3) model and the QM fitted the experimental data of WRHDs except for the latest BESIII measurement.  An interesting observation is that although for the $\Lambda\to n\gamma$ decay the broken SU(3) model and the QM  differ on the branching ratio, they predicted a similar asymmetry parameter, which is away from the measured value by more than $5\sigma$. In terms of the signs of the asymmetry parameters, the broken SU(3) model, the QM, and the EOMS $\chi$PT did a similar work. The EOMS $\chi$PT  can reproduce four of the five measured modes. The only discrepancy is in the $\Sigma^+\to p \gamma$ decay,  where the predictions are different from the PDG averages and the latest BESIII data~\cite{BESIII:2023fhs} by almost $5\sigma$, considering simultaneously the branching ratio and decay parameter. Given the fact that EOMS $\chi$PT is a model independent approach, such a discrepancy calls for more theoretical and experimental scrutinise of this particular channel. We note that in the EOMS $\chi$PT~\cite{Shi:2022dhw}, only the contributions of intermediate octet baryons are considered, while those of heavier resonances, such as the $\Lambda(1405)$ state,  are not, which may play a non-negligible role as pointed out in Refs.~\cite{Borasoy:1999nt,Zenczykowski:1999vq} and need to be studied in the future.  In addition, the EOMS $\chi$PT results~\cite{Shi:2022dhw} suffer from the so-called $S/P$ puzzle, which leads to the fact that only correlations between branching ratios and asymmetry parameters can be given. We expect to see more theoretical studies along these directions.
\begin{table*}[htb!]
\centering
\caption{\label{tab1} Experimental data~\cite{E761:1993unn,BESIII:2023fhs,ParticleDataGroup:2022pth,BESIII:2022rgl} and theoretical predictions of branching fractions ${\cal B}$ and asymmetry parameters $\alpha_\gamma$ of the WRHDs.}
{\small
  \begin{tabular}{ccccc}
\hline
\hline
 ~~~Decay modes~~~ & ~~~Data~\cite{E761:1993unn,BESIII:2023fhs,ParticleDataGroup:2022pth,BESIII:2022rgl}~~~ & ~~~Broken SU(3) model~\cite{Zenczykowski:2005cs}~~~ & ~~~QM~\cite{Dubovik:2008zz}~~~ & ~~~NRCQM~\cite{Niu:2020aoz}~~~\\
\hline
 & \multicolumn{4}{c}{${\cal B}\times10^{-3}$}\\
 \cline{2-5}
$\Lambda\to n\gamma$ & $0.832(38)(54)$ & $0.77$ & $1.84$ & 1.83(96)\\

$\Sigma^+\to p\gamma$ & $0.996(21)(18)$  & $0.72$ & $1.30$ & $1.06(59)$\\

$\Sigma^0\to n\gamma$ & $\cdots$ & $\cdots$ & $4.3\times10^{-9}$ & $10^{-10}$\\

$\Xi^0\to\Lambda\gamma$ & $1.17(7)$ & $1.02$ & $0.93$ & $0.96(32)$\\

$\Xi^0\to\Sigma^0\gamma$ & $3.33(10)$ & $4.42$ & $3.82$ & $9.75(418)$\\

$\Xi^-\to \Sigma^-\gamma$ & $0.127(23)$ & $0.16$ & $0.13$ & $\cdots$\\
\cline{2-5}
 & \multicolumn{4}{c}{$\alpha_\gamma$}\\
 \cline{2-5}
$\Lambda\to n\gamma$  & $-0.16(10)(5)$ & $-0.93$ & $-0.94$ & $-0.67(6)$\\

$\Sigma^+\to p\gamma$ & $-0.652(56)(20)$ & $-0.67$ & $-0.74$ & $-0.58(6)$\\

$\Sigma^0\to n\gamma$  & $\cdots$ & $\cdots$ & $0.01$ & $0.37(4)$\\

$\Xi^0\to\Lambda\gamma$ & $-0.704(19)(64)$ & $-0.97$ & $-0.64$ & $0.72(11)$\\

$\Xi^0\to\Sigma^0\gamma$  & $-0.69(6)$ & $-0.92$ & $-0.52$ & $0.33(4)$\\

$\Xi^-\to \Sigma^-\gamma$  & $1.0(13)$ & $0.8$ & $0.76$ & $\cdots$\\
\hline
\hline
\end{tabular}
}
\end{table*}

Note that the Lagrangians of the EOMS $\chi$PT are local and satisfy the CPS symmetry, which is CP followed by the SU(3) transformation of $u \rightarrow-u, d \rightarrow s, s \rightarrow d$ which exchanges $\mathrm{s}$ and $\mathrm{d}$ quarks, and therefore predicts that the asymmetry parameters of $\Sigma^+\to p\gamma$ and $\Xi^-\to\Sigma^-\gamma$ are vanishing in the exact U-spin symmetric limit, which satisfies Hara's theorem. The non-vanishing values of $\alpha_\gamma(\Sigma^+\to p\gamma)$ and $\alpha_\gamma(\Xi^-\to \Sigma^-\gamma)$ are due to the breaking of U-spin symmetry. We must stress that Hara's theorem only applies to the U-spin symmetric limit, while in reality because of the mass difference between the down quark and the strange quark, it must be violated. As a result its violation is expected. The puzzle is the magnitude of its violation in particular in the $\Sigma^+\to p\gamma$ decay. The studies of models in Refs.~\cite{Zenczykowski:2005cs,Dubovik:2008zz,Niu:2020aoz} and EOMS $\chi$PT~\cite{Shi:2022dhw} all indicate that the violation of Hara's theorem is mainly due to SU(3)-flavor (U-spin) symmetry breaking. In this regard, $\chi$PT is a more useful framework because it follows a self-consistent power-counting and allows one to study SU(3)-flavor symmetry breaking in a systematic and model independent way~\cite{Geng:2013xn}. From this perspective, the discrepancy in the $\Sigma^+\to p\gamma$ channel is indeed a bit disturbing given that the latest BESIII data~\cite{BESIII:2023fhs} largely agree with the current PDG data but more precise. We further notice that as discussed in detail in the EOMS $\chi$PT work~\cite{Shi:2022dhw}, hyperon non-leptonic decays are relevant to understanding the WRHDs. Considering that the new BESIII measurements~\cite{BESIII:2021ypr,BESIII:2018cnd,BESIII:2022qax} of asymmetry parameters of $\Lambda$ and $\Xi^-$ non-leptonic decays are different from the previous ones by more than $5\sigma$ and almost $2\sigma$, respectively, we encourage the BESIII Collaboration to reinvestigate the non-leptonic decays of other hyperons, i.e., $\Sigma^+$ and $\Xi^0$, in the future in order to better understand 
the puzzle of WRHDs. In addition, as already stressed in Ref.~\cite{Shi:2022dhw}, the current $\alpha_\gamma(\Xi^-\to\Sigma^-\gamma)$ has a rather large uncertainty but the EMOS $\chi$PT made a more precise prediction, which differs from the experimental central value, as well as the predictions of the broken SU(3) model and the QM. Therefore, it will be interesting to see whether a more precise measurement of the asymmetry parameter $\alpha_\gamma(\Xi^-\to\Sigma^-\gamma)$ agrees with the prediction of the EOMS $\chi$PT~\cite{Shi:2022dhw}, i.e., $\alpha_\gamma(\Xi^-\to\Sigma^-\gamma)\in[-0.18,~0.38]$, assuming that the branching ratio remains unchanged. In the future, if this is indeed the case, the measured result will not only be able to distinguish between different models but also can help better understand the current experimental value of $\alpha_\gamma(\Sigma^+\to p\gamma)$ and shed light on the nature of WRHDs, because they are strong correlated in baryon chiral perturbation theory. It is therefore worth performing such a measurement in the future.

In the proposed super tau-charm factory~\cite{Zhou:2021rgi,Barnyakov:2020vob}, an electron-positron collider where the expected peak luminosity and center-of-mass energy are $(0.5\sim1.0)\times10^{35}~{\rm cm}^{-2}{\rm s}^{-1}$ and $2\sim7~{\rm GeV}$, about $3.4\times10^{12}$ $J/\psi$ events can be collected with a running period of one year. Even assuming the same experimental technique, the increase of statistics alone can reduce the statistical uncertainty by at least one order of magnitude. The anticipated results in the super tau-charm factory will undoubtly deepen our understanding on the puzzle of WRHDs. In addition, more precise measurements of WRHDs can provide essential standard model inputs to studies of new physics beyond the standard model in rare semi-leptonic hyperon decays~\cite{Geng:2021fog,He:2018yzu} and are relevant to parity violating photoproduction of $\pi^\pm$ on the $\Delta(1232)$ resonance~\cite{Zhu:2001br}. 

\section*{Conflict of interest}

The authors declare that they have no conflict of interest.

\section*{Acknowledgments}
We thank Dr. Qiang Zhao and Dr. Peng-Yu Niu for enlightening discussions about the NRCQM results. This work is supported in part by the National Natural Science Foundation of China under Grants No.11735003, No.11975041,  and No.11961141004. RXS acknowledges support from the Project funded by China Postdoctoral Science Foundation No.2021M700343.

\bibliographystyle{elsarticle-num}
\bibliography{WRHDs}


\end{document}